\documentclass[aps,prl,superscriptaddress,10pt,twocolumn]{revtex4-2}

\usepackage[T1]{fontenc}
\usepackage[utf8]{inputenc}
\usepackage{amsmath}
\usepackage{amssymb}
\usepackage{graphicx}
\usepackage{chemformula}
\usepackage{siunitx}
\usepackage[colorlinks=true, allcolors=blue]{hyperref}
\usepackage{multirow}

\usepackage{tabularx}

\begin{document}

\author{Ismail Eren}
\affiliation{Centrum for Advanced Systems Understanding, CASUS, Conrad-Schiedt-Str. 20, 02826 Görlitz, Germany}
\affiliation{Helmholtz-Zentrum Dresden-Rossendorf, Bautzner Landstr. 400, 01328 Dresden, Germany}

\author{Ege Yigit Erbil}
\affiliation{Koc Üniversitesi, Rumelifeneri Yolu 34450 Sarıyer, Istanbul, Türkiye}

\author{Maria-Judith Caisachana-Lozada}
\author{Hossein Mirhosseini}
\author{Thomas D. K\"uhne}
\author{Agnieszka B. Kuc}
\affiliation{Centrum for Advanced Systems Understanding, CASUS, Conrad-Schiedt-Str. 20, 02826 Görlitz, Germany}
\affiliation{Helmholtz-Zentrum Dresden-Rossendorf, Bautzner Landstr. 400, 01328 Dresden, Germany}
\email{a.kuc@hzdr.de}

\title{Capturing Nuclear Quantum Effects in Hydrogen Diffusion inside MoS$_2$ via Machine-Learning-Enhanced Path-Integral Simulations}

\maketitle

\date{\today}

\section{Abstract}
Hydrogen transport through layered two-dimensional (2D) materials is central to technologies such as hydrogen storage, fuel cells, and isotope separation. Among these materials, MoS$_2$ exhibits tunable interlayer diffusion properties, whose accurate theoretical description requires accounting for nuclear quantum effects (NQEs), including zero-point motion and tunneling.
Here, we present a machine-learning-enhanced atomistic study of hydrogen and deuterium diffusion in layered MoS$_2$ based on interatomic potentials trained on r$^2$SCAN+rVV10 density-functional-theory data. Combining well-tempered metadynamics with path-integral molecular dynamics, we investigate diffusion across multiple MoS$_2$ polytypes and twisted bilayer structures while explicitly incorporating NQEs.
Our simulations show that NQEs substantially lower free-energy barriers for hydrogen diffusion at 300 K, significantly increasing the hydrogen self-diffusion coefficient compared to classical nuclei simulations. We further identify a pronounced kinetic isotope effect, with a 31 meV difference between hydrogen and deuterium quantum free-energy barriers. In twisted bilayer MoS$_2$, hydrogen transport exhibits strong spatial variations governed by the local stacking environments within the moiré superlattices.
These results highlight the critical role of NQEs in hydrogen transport through layered materials and provide atomistic insight into isotope-selective diffusion in structurally complex 2D systems.

\section{Introduction}
Layered two-dimensional (2D) van der Waals (vdW) materials have long attracted attention due to their ability to host intercalated atomic and ionic species within their confined interlayer regions.
Such intercalation processes, involving ions such as Li$^+$ and Na$^+$, are considered for numerous energy-related technologies.\cite{Fan2017, Ji2022, Shao2023, Astles2024, Liers2025}
Owing to its exceptionally small size, hydrogen can also penetrate these vdW gaps, making layered materials promising platforms for hydrogen transport and isotope separation.
In particular, experimental studies on materials such as $h$-BN and MoS$_2$ have demonstrated interlayer hydrogen diffusion together with isotope-selective transport arising from different entry barriers for protium and deuterium.\cite{hu2018transport}

Beyond their intercalation and transport properties, layered 2D materials offer exceptional tunability through layer engineering.
In particular, stacking arrangements and relative twist angles introduce additional structural degrees of freedom that enable the tailoring of electronic, optical, and transport properties for next-generation applications.\cite{cao2018unconventional,arnold2023,ramzan2023effect,yuan2020twist,ahn2018dirac,ni2019soliton}

In our previous studies, we have shown that hydrogen diffusion in transition-metal dichalcogenides (TMDCs), including different high-symmetry MoS$_2$ stackings, strongly depends on the local interlayer environment.\cite{eren2024hydrogen,an2019chemistry}
However, most theoretical investigations have relied on the classical treatment of nuclei, despite the fact that light atoms exhibit pronounced nuclear quantum effects (NQEs), even at room temperature.\cite{ege8,ege9,ege10}
These effects, including quantum tunneling, zero-point energy (ZPE), and quantum delocalization, can be incorporated through path-integral molecular dynamics (PIMD).\cite{ege15}
Our previous work further demonstrated that hydrogen diffusivity is highly sensitive to the relative layer orientation, suggesting twisted bilayers as an attractive platform for studying hydrogen transport.\cite{eren2024hydrogen} Twisted 2D materials are known to host rich moiré-driven phenomena, including flat-band superconductivity, soliton formation, and spatially modulated excitonic transport.\cite{cao2018unconventional,arnold2023,ramzan2023effect,yuan2020twist} Analogous to the way moiré potentials govern electrons and excitons, the structural relaxations and local stacking variations in twisted bilayers are expected to create complex diffusion landscapes for hydrogen migration.
Accurately describing such systems remains computationally challenging. While density functional theory (DFT) provides an accurate description of the electronic structure,\cite{ege1,ege2,ege3} advanced exchange-correlation functionals required to capture both vdW interactions and hydrogen bonding effects, such as r$^2$SCAN+rVV10, substantially increase the computational cost.\cite{ege4,ege5,ege6}
The challenge becomes even more severe when combining first-principles calculations with PIMD and enhanced-sampling techniques for large twisted supercells.\cite{ege7,ege10}

Machine-learning interatomic potentials (MLIPs) have recently emerged as an effective approach to bridge this gap.\cite{ege13}
Trained on high-fidelity DFT datasets, MLIPs can reproduce near-DFT accuracy while enabling simulations over substantially larger length and time scales.\cite{ege7,ege14,ege16,ege17}
This acceleration is particularly important for studies involving quantum nuclear dynamics and enhanced configurational sampling.\cite{ege18,ege16,ege19}

In this work, we investigate the role of NQEs in hydrogen and deuterium diffusion across MoS$_2$ polytypes and selected twisted bilayer structures using a robust MLIP trained on r$^2$SCAN+rVV10 DFT data.
First, we demonstrate that NQEs significantly influence hydrogen diffusion in all high-symmetry MoS$_2$ stackings, with the strongest effects (difference of about 58 meV) observed for the $H_h^h$ configuration, which is the most stable polytype of this material.
Second, we reveal a pronounced isotope effect, where the quantum contribution to the free-energy barrier is approximately 31 meV smaller for deuterium than for protium.
Finally, we study hydrogen diffusion in twisted bilayers with twist angles of $\theta = 3.89^\circ$ and $21.79^\circ$ ($R$ stacking), and $38.21^\circ$ and $56.11^\circ$ ($H$), showing that diffusion within the staggered regimes is preferential over the eclipsed ones, owing to the favorable proximity of chalcogen atoms in the neighboring layers.
By combining MLIP-driven simulations with path-integral methods and complex twisted structures, this work establishes a comprehensive atomistic framework for understanding how quantum fluctuations and interlayer twisting cooperatively dictate hydrogen transport in 2D TMDC materials.

\section{Methodology}

\begin{figure*}[ht!]
\centering
 \includegraphics[width=1\textwidth]{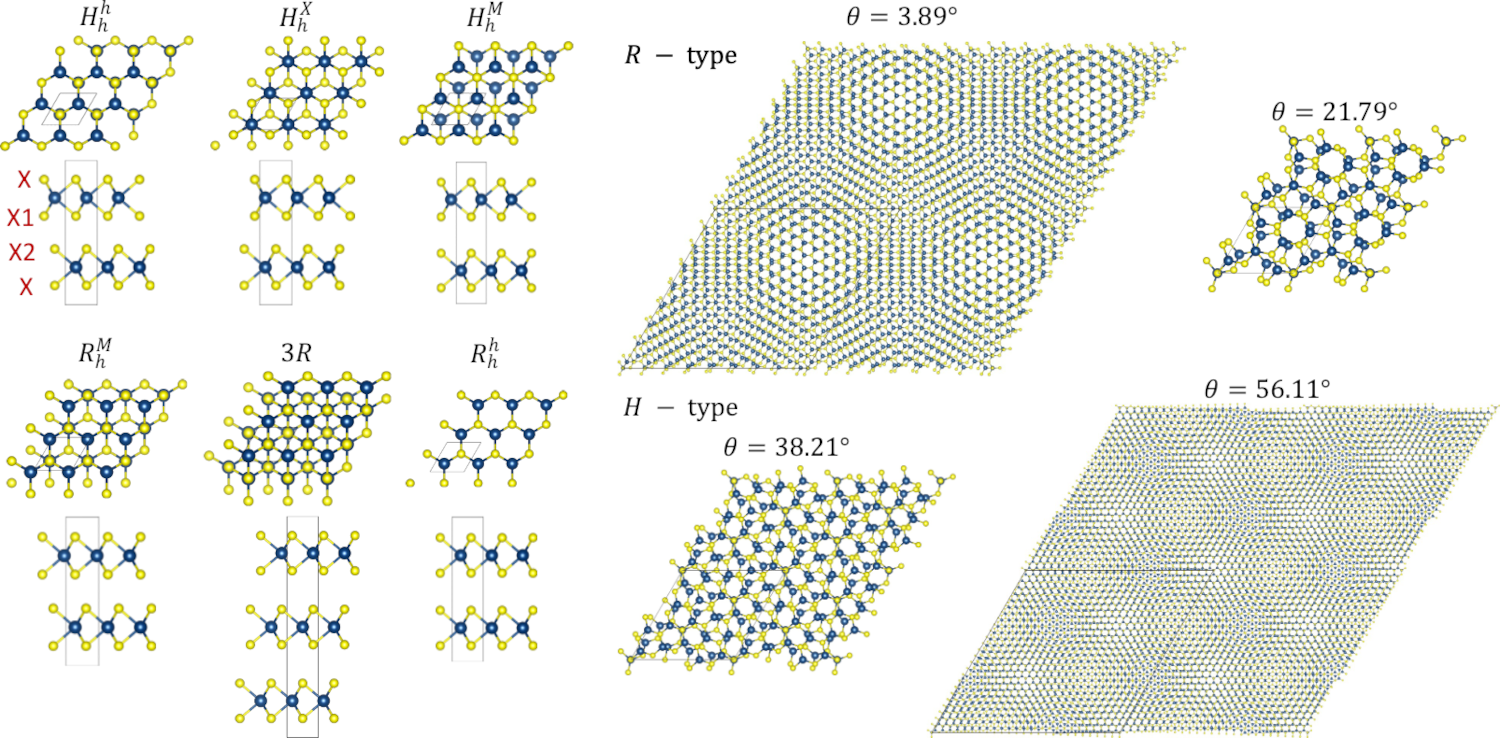}%
 \caption{\label{fig:Structures} 
Atomic structures of high-symmetry stackings of MoS$_2$ and selected twisted bilayers (as optimized by Arnold et al.~\cite{arnold2023}) shown in supercell representation. Yellow: S atoms, blue: Mo atoms. The collective variables are based on the following division of S atoms: X$_1$ and X$_2$ denote the S atoms that bind to H atom, X are all the other S atoms in the system. For details, see Methodology Section.
 }
\end{figure*}

We investigated hydrogen diffusion within the interlayer space of bulk and bilayer MoS$_2$ systems, considering several high-symmetry stacking configurations, including $H^{h}_{h}$, $H^{X}_{h}$, $R^{M}_{h}$, and 3R phases (see Fig.~\ref{fig:Structures}).
In addition, isotope effects were examined for bulk $H^{h}_{h}$-MoS$_2$, while hydrogen transport in twisted bilayers was studied for representative twist angles of $\theta = 3.89^\circ$, $21.79^\circ$, $38.21^\circ$, and $56.11^\circ$.

To efficiently access the large length and time scales required for quantum diffusion simulations, we developed a machine-learning interatomic potential (MLIP) based on the MACE framework.\cite{batatia2022mace}
The potential was trained on a compact density-functional-theory dataset (see the Supporting Information for details) and subsequently employed in both Born-Oppenheimer molecular dynamics (which treats nuclei classically) and path-integral molecular dynamics (PIMD; which treats nuclei quantum-mechanically) simulations.\cite{marx1994ab,barducci2008well}
Nuclear quantum effects were incorporated using the i-PI package,\cite{Ceriotti2016} while free-energy barriers ($\Delta F$) for interlayer hydrogen diffusion were evaluated through well-tempered metadynamics (WTMetaD) implemented in PLUMED.\cite{barducci2008well,plumed1,plumed2}

All dynamical simulations were performed using the i-PI package.\cite{Ceriotti2016}
Classical nuclei simulations were treated within the single-bead limit ($P=1$), corresponding to standard molecular dynamics.
Simulations were carried out in the NVT ensemble using a time step of 0.1~fs and a stochastic velocity rescaling (SVR) thermostat.

To account for nuclear quantum effects, PIMD simulations were performed using multiple beads to represent each quantum nucleus.
The required number of beads, $P$, was estimated from
\begin{equation}\label{eq1}
P \geq \frac{\hbar \omega_{\mathrm{max}}}{k_{B}T},
\end{equation}
where $\hbar$ is the reduced Planck constant, $k_{B}$ is the Boltzmann constant, $T$ is the temperature, fixed here at 300~K, and $\omega_{\mathrm{max}}$ corresponds to the highest vibrational frequency of the S--H bond.
Substituting the corresponding values yields $P \geq 12.35$; therefore, $P=16$ beads were employed throughout this work to ensure numerical convergence and computational efficiency.
Additional convergence tests are provided in the Supporting Information. The PIMD simulations were performed within the ring-polymer molecular dynamics (RPMD) framework using the global path-integral Langevin equation thermostat (PILE-G).

Free-energy barriers for hydrogen and deuterium interlayer diffusion, $\Delta F$, were evaluated using WTMetaD and subsequently employed to estimate the self-diffusion coefficients.\cite{beyer1982determination, eren2024hydrogen, an2019chemistry}
Gaussian hills with an initial height of 25~meV and a width of 0.1~\AA\ were deposited every 1000 simulation steps.
For computational efficiency, the metadynamics bias was applied only to the centroid coordinate in the PIMD simulations.

The S--H stretching frequency was taken as 2570~cm$^{-1}$ from the literature,\cite{freqs} assuming negligible contributions from second-neighbor interactions.
Since hydrogen and deuterium possess identical electronic structures but different masses, the vibrational frequencies were related through
\begin{equation}\label{eq2}
\nu = \frac{1}{2\pi c}\sqrt{\frac{k}{\mu}},
\end{equation}
where $\mu$ is the reduced mass and $k$ is the effective bond force constant.
Using the corresponding mass ratio, the S--D stretching frequency was estimated to be 1845~cm$^{-1}$.

The collective variables used in WTMetaD were defined through the coordination number between H(D) and neighboring S atoms.\cite{eren2024hydrogen, an2019chemistry}
A cutoff distance of $R_0 = 1.6$~\AA\ and a switching parameter of $d_0 = 0.2$~\AA\ were employed.
Prior to each WTMetaD production run, the systems were equilibrated for at least 1~ps using a time step of 0.1~fs to ensure thermal equilibrium, see Fig.~S4 in the Supporting Information.

\section{Results}
Before performing the diffusion simulations, we validated the trained MLIP by examining the energetic stability and phonon band structures of the investigated high-symmetry MoS$_2$ stackings.
The MLIP results were compared against the corresponding DFT reference calculations and showed relatively good agreement for both relative energetics and vibrational properties, confirming the reliability of the potential for describing the structural and dynamical behavior of the system.
Detailed comparisons are provided in the Supporting Information Figs.~S2 and S3.

Next, building upon our previous work,\cite{eren2024hydrogen} we first reproduced the classical-nuclei WTMetaD calculations using the developed MLIP framework.
Subsequently, we combined PIMD with WTMetaD to investigate how NQEs influence hydrogen diffusion within the interlayer regions of high-symmetry bulk MoS$_2$ phases.
The structural and dynamical accuracy of the MLIP is discussed in the Methodology Section and in the Supporting Information.

\begin{figure}[ht!]
\centering
 \includegraphics[width=0.35\textwidth]{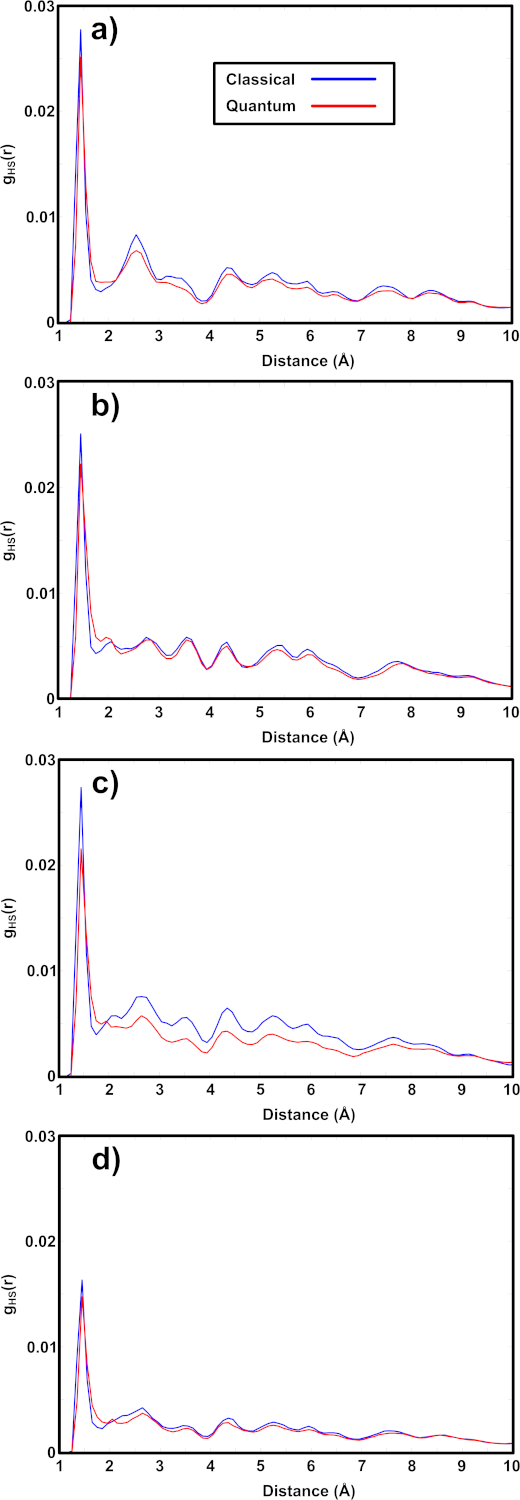}%
 \caption{\label{fig2} 
Radial pair distribution functions, $g_{HS}(r)$, for H and S atoms in selected MoS$_2$ stacking types: (a) $H_{h}^{h}$, (b) $H_{h}^{X}$, (c) $R_{h}^{M}$, and (d) $3R$. Results are obtained from MLIP-based molecular dynamics using classical nuclei (blue, $P=1$) and quantum nuclei via PIMD (red, $P=16$). The two high-energy stackings, $H_{h}^{M}$ and $R_{h}^{h}$, are not shown because they shift to more favorable layer arrangements.
 }
\end{figure}

The structural correlations between H and S atoms are characterized using the radial distribution functions (RDFs), $g_{HS}(r)$, shown in Fig.~\ref{fig2}.
For the PIMD simulations, the centroid positions of the beads were used to represent the quantum nuclei.
All investigated phases exhibit a pronounced first peak at approximately 1.45~\AA, corresponding to the primary H--S bond.
Compared to the classical treatment ($P=1$), the quantum simulations ($P=16$) display slightly broadened and less intense peaks, reflecting the increased spatial delocalization induced by zero-point motion and quantum fluctuations. 
This effect is particularly pronounced for the first coordination shell, where H atoms sample a broader region of the free-energy surface.

\begin{figure*}[ht!]
\centering
 \includegraphics[width=1.0\textwidth]{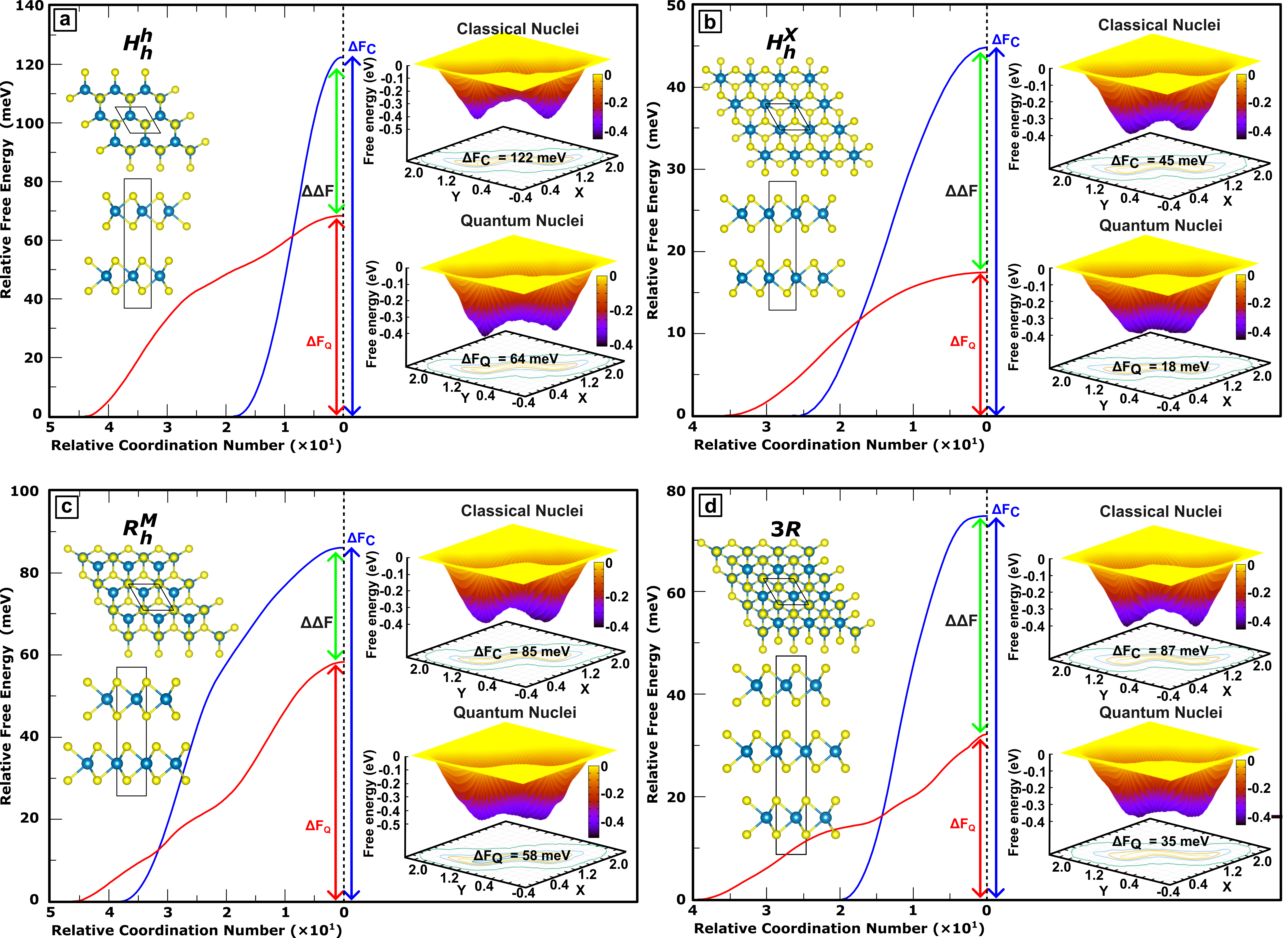}%
 \caption{\label{fig3} 
 Free-energy surfaces (FES) for hydrogen diffusion in selected MoS$_2$ stacking types: (a) $H_h^h$, (b) $H_h^X$, (c) $R_h^M$, and (d) 3R MoS$_2$. Each panel displays the top and side views of the respective atomic structure. The curves on the left illustrate the minimum-energy Pathway (MEP) of relative free energy (in meV) as a function of the relative coordination number (RCN) up to the barrier top. These were calculated using WTMetaD. Although WTMetaD employs two collective variables, the RCN is derived from CV1, with the barrier top shifted to a CN of 0. The free energy is relative to the MEP minimum (set to 0) to highlight the diffusion barriers. Solid blue lines represent the classical free-energy surface  MEP, while solid red lines correspond to the Path Integral Molecular Dynamics (PIMD) MEP. Key energy barriers are indicated by $\Delta F_C$ (classical free energy barrier), $\Delta F_Q$ (quantum free energy barrier), and $\Delta \Delta F$ (the difference between classical and quantum barriers). The right side of each panel features the raw 3D FES colormaps for both classical and quantum nuclei calculations, annotated with their respective diffusion barrier values.
 }
\end{figure*}

Beyond the primary H--S interaction, the RDFs reveal distinct structural environments depending on the stacking configuration.
The $H_{h}^{h}$ and $H_{h}^{X}$ phases [Fig.~\ref{fig2}(a,b)] exhibit relatively well-defined higher-order peaks, indicating a more ordered hydrogen environment within the interlayer space.
In contrast, the $R_{h}^{M}$ stacking [Fig.~\ref{fig2}(c)] shows pronounced quantum-induced broadening even at larger distances, demonstrating that NQEs remain significant beyond the nearest-neighbor coordination shell.
The 3R phase [Fig.~\ref{fig2}(d)] displays substantially weaker long-range correlations, suggesting a more disordered and dynamically flexible hydrogen environment.

Overall, the inclusion of NQEs systematically reduces structural ordering across all investigated phases, indicating that classical nuclei simulations tend to over-localize hydrogen within the interlayer region.
These results demonstrate that an accurate description of hydrogen transport in layered TMDCs requires explicit treatment of the quantum nature of light nuclei.

Next, we evaluated the free-energy surfaces (FESs) of the investigated MoS$_2$ phases using both classical and quantum treatments of the nuclei.
Fig.~\ref{fig3} shows the resulting FESs together with the corresponding minimum-energy diffusion pathways for hydrogen migration within the interlayer region.
First, we observe that the developed MLIP accurately reproduces the trends previously obtained from first-principles WTMetaD simulations,\cite{eren2024hydrogen} confirming its reliability for describing hydrogen diffusion in layered MoS$_2$.
In particular, the MLIP captures the systematically lower $\Delta F$ values associated with the $R$-type stackings compared to the $H$-type configurations.
The only exception is observed for the $H^X_h$ stacking, for which we underestimate $\Delta F$.
Second, a clear reduction in $\Delta F$ is observed upon inclusion of NQEs, even at 300~K.
Across all investigated polytypes, the quantum treatment leads to a noticeable flattening and softening of the free-energy landscape relative to the classical description.
As a consequence, the diffusion pathways become energetically more accessible, resulting in substantially lower $\Delta F$ for hydrogen transport.
These results demonstrate that zero-point motion and quantum delocalization significantly modify the effective diffusion landscape in layered MoS$_2$, highlighting the importance of explicitly accounting for NQEs when describing hydrogen migration in 2D materials.

A quantitative analysis of the free-energy surfaces reveals that classical simulations systematically overestimate the barriers for hydrogen diffusion in all investigated MoS$_2$ phases.
The inclusion of NQEs substantially lowers the effective diffusion barriers, with reductions ranging from tens of meV up to nearly 58~meV depending on the stacking configuration (Fig.~\ref{fig3}).
The strongest quantum correction is observed for the $H_h^h$ phase, the energetically most favorable stacking, where the barrier decreases from 122~meV in the classical description to only 64~meV in the quantum treatment. 
In contrast, the $H_h^X$ configuration exhibits the lowest quantum $\Delta F$ overall, decreasing from 45~meV to just 18~meV. However, the absolute barrier values for this configuration are underestimated.
Significant barrier reductions are also found for the $R_h^M$ and 3R phases, whose barriers decrease from 85 and 87~meV down to 58 and 35~meV, respectively.
These results demonstrate that NQEs not only soften the diffusion landscape but can also substantially alter the relative diffusion efficiencies between different stacking configurations.

\begin{figure}[ht!]
\centering
 \includegraphics[width=0.6\columnwidth]{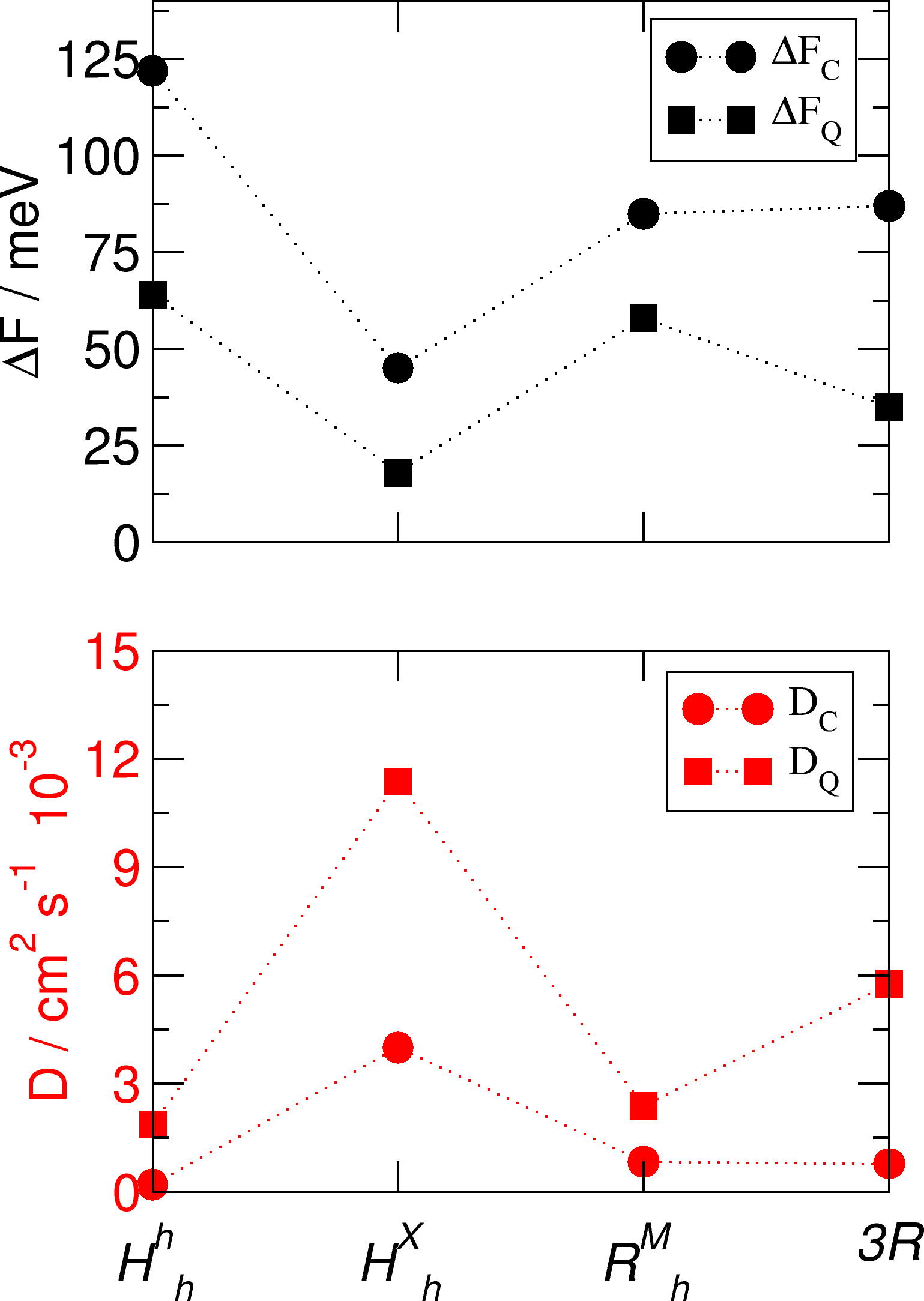}%
 \caption{\label{fig4} 
Free energy barriers ($\Delta F$; black) and the corresponding self-diffusion coefficients (D; red) of MoS$_2$ high-symmetry stackings with classical (C) and quantum (Q) treatment of the nuclei.
 }
\end{figure}

Fig.~\ref{fig4} summarizes the classical and quantum free-energy barriers together with the corresponding $D$.
Consistent with the barrier reductions observed in the quantum free-energy surfaces, the inclusion of NQEs leads to a pronounced enhancement of hydrogen diffusivity across all investigated MoS$_2$ phases.
Among the considered stackings, the $H^{X}_{h}$ configuration exhibits the highest diffusion coefficient, reaching $11.4 \times 10^{-3}$ cm$^2$s$^{-1}$ in the quantum simulations.
In contrast, the $H^{h}_{h}$ phase remains the system with the lowest diffusivity within the classical description, but also displays the strongest sensitivity to NQEs.
Here, $D$ increases by nearly an order of magnitude, corresponding to a quantum correction factor (QCF) of 9.43.
Similarly, the 3R phase shows a substantial quantum enhancement, with a QCF of 7.47.
The $H^{X}_{h}$ and $R^{M}_{h}$ stackings exhibit comparatively smaller, yet still significant, quantum corrections, both with QCF values close to 2.84.
These results demonstrate that the impact of NQEs strongly depends on the local stacking environment and the corresponding diffusion landscape.
In particular, phases characterized by larger classical diffusion barriers experience the most pronounced quantum enhancement of hydrogen transport.

These results further confirm that interlayer stacking strongly governs hydrogen transport in MoS$_2$, both within classical and quantum descriptions.
While the inclusion of NQEs systematically enhances hydrogen diffusivity, the relative diffusion trends between the different stackings remain largely preserved.
In particular, the $H_h^X$ configuration exhibits the highest diffusivity in both the classical and quantum regimes, consistent with its exceptionally low free-energy barriers. However, the absolute values are underestimated in this case.
In contrast, although the $H_h^h$ phase experiences the largest quantum-induced barrier reduction, it still retains the lowest quantum diffusion coefficient, indicating that the overall topology of the diffusion landscape remains less favorable for hydrogen migration.
This highlights that the magnitude of the quantum correction alone does not directly determine the absolute diffusivity, which instead emerges from the combined interplay between barrier heights and the broader structural environment.

Interestingly, the Bernal-stacked R-type phases exhibit similar diffusion coefficients within the classical treatment despite their distinct stacking sequences (ABAB vs. ABC).
However, noticeable differences emerge in the quantum regime, where the 3R phase displays substantially enhanced diffusivity compared to $R_h^M$.
This behavior suggests that neighboring-layer arrangements and the interlayer interactions play a more pronounced role once quantum delocalization is included, emphasizing the sensitivity of hydrogen transport to subtle structural variations in layered MoS$_2$.

\begin{figure}[ht!]
\centering
 \includegraphics[width=0.8\columnwidth]{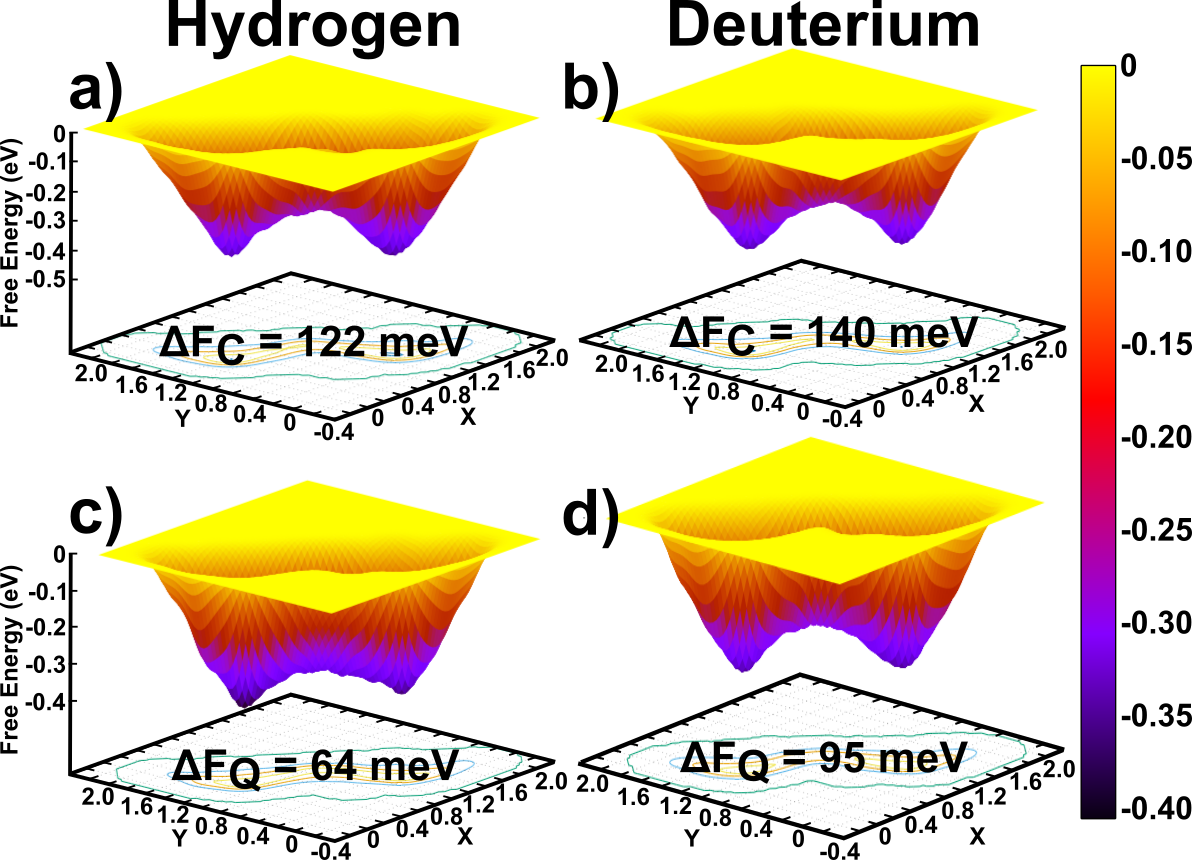}%
 \caption{\label{fig5} 
Comparison of classical (a, b) and quantum (c, d) treatment of the nuclei during hydrogen (a, c) and deuterium (b, d) diffusion between layers of $H^h_h$-MoS$_2$. The corresponding free energy barriers, $\Delta F_C$ and $\Delta F_Q$ are also given.
 }
\end{figure}

Next, we investigated the isotope effects in hydrogen diffusion by explicitly incorporating NQEs through RPMD.
While isotope-selective transport in layered materials has previously been explored experimentally by Hu et al.,\cite{hu2018transport} those studies primarily focused on entry barriers at the material interface.
As a complementary approach, our simulations provide direct atomistic insight into isotope-dependent interlayer diffusion within bulk MoS$_2$.
Using the same computational framework as for protium, we calculated the FESs for deuterium diffusion in $H_h^h$-MoS$_2$ (Fig.~\ref{fig5}), while the resulting $\Delta F$ and $D$ are summarized in Table~\ref{tab:table4}.
Owing to the larger mass of deuterium, the corresponding S--D vibrational frequency is reduced, leading to a lower bead requirement for PIMD simulations ($P \geq 8.8$ according to Eq.~\ref{eq1}).
Therefore, the choice of $P=16$ beads, already used for protium, is sufficient for converged deuterium simulations.

\begin{table}[ht!]
\centering
 \caption{\label{tab:table4} The free-energy barriers, $\Delta F$ and the corresponding self-diffusion coefficients, $D$, for protium and deuterium diffusion between layers of $H_h^h$-MoS$_2$ calculated with classical and quantum treatment of the nuclei. 
 The energy differences between the two isotopes indicate a strong kinetic isotope effect.}
 \begin{tabular}{cccc}
 \hline
  Particle & Nuclei & $\Delta F$ & $D$ \\
     &  & meV & cm$^2$s$^{-1}$ $\times10^{-3}$ \\
  \hline
  H & Classical & 122 & 0.198 \\
  H & Quantum & 64 & 1.870 \\
  D & Classical & 140 & 0.071 \\
  D & Quantum & 95 & 0.404 \\
  \hline
 \end{tabular}
\end{table}

The calculated FESs reveal a pronounced isotope dependence of the diffusion barriers.
For protium, the inclusion of NQEs reduces the $\Delta F$ from 122~meV to 64~meV, corresponding to a 48\% reduction.
In contrast, deuterium exhibits a smaller quantum reduction, with the barrier decreasing from 140~meV to 95~meV, corresponding to a reduction of only 32\%.
This weaker quantum softening directly reflects the larger nuclear mass and weaker quantum character of deuterium, demonstrating the strong mass dependence of quantum diffusion in layered materials.
The resulting 31 meV difference between the quantum free-energy barriers of protium and deuterium also indicates a pronounced kinetic isotope effect (KIE), even at 300~K.

Interestingly, isotope-dependent differences are already present within the classical description, where deuterium exhibits systematically larger diffusion barriers and lower diffusivity than hydrogen.
However, these differences become substantially amplified once NQEs are included.
While the diffusion coefficient of deuterium increases under the quantum treatment, it remains significantly lower than that of protium, leading to an enhanced isotope selectivity in the quantum regime.

Overall, these results demonstrate that NQEs play a decisive role not only in enhancing hydrogen transport, but also in governing isotope-selective diffusion in layered MoS$_2$.
The explicit inclusion of quantum nuclear dynamics is therefore essential for accurately describing hydrogen isotope transport in 2D materials.

In our previous work,\cite{eren2024hydrogen} as well as in the preceding sections of this study, we demonstrated that hydrogen diffusivity strongly depends on the local high-symmetry stacking environment in MoS$_2$.
Since these stackings naturally emerge within twisted bilayer MoS$_2$ (TBL-MoS$_2$) through moiré reconstruction,\cite{arnold2023,ni2019soliton,ahn2018dirac} twisted systems provide an ideal platform for investigating spatially varying hydrogen transport.
We therefore extended our analysis to TBL-MoS$_2$ structures with twist angles of $\theta = 3.89^\circ$, $21.79^\circ$, $38.21^\circ$, and $56.11^\circ$, which we will refer to as TBL-MoS$_2^{\theta}$.

For the moiré regime without pronounced domain formation ($\theta = 21.79^\circ$, $R$-type stackings and $\theta = 38.21^\circ$, $H$-type stackings), PIMD simulations reveal that nuclear quantum effects remain significant, lowering $\Delta F$ by approximately 40~meV and 80~meV relative to the classical description, for 21.79$^\circ$ and 38.21$^\circ$, respectively.
Interestingly, both systems exhibit nearly identical quantum diffusion barriers of approximately 120~meV (see Tab.~\ref{tab:table5}), suggesting that hydrogen transport becomes progressively less sensitive to the precise twist angle as the system approaches the quasicrystalline regime near $\theta = 30^\circ$.\cite{ahn2018dirac}
This behavior originates from the absence of extended high-symmetry domains in these large-angle moiré structures.
Instead, the local stacking landscape is dominated by continuously varying intermediate configurations, while energetically favorable high-symmetry regions occupy only very small areas.
As a result, hydrogen diffusion is governed primarily by these intermediate stacking environments, leading to substantially larger $\Delta F$ values than those of isolated high-symmetry stackings and, consequently, diffusion coefficients that are at least one order of magnitude lower.

\begin{table}[ht!]
\centering
 \caption{\label{tab:table5} Hydrogen diffusion in selected twisted bilayer MoS$_2$. Twist angles, $\theta$, free energy barriers, $\Delta F$, and the corresponding self-diffusion coefficients, $D$ calculated with classical and quantum treatment of the nuclei.
 }
 \begin{tabular}{cccc}
 \hline
  $\theta$  & Nuclei & $\Delta F$ & $D$ \\
    $^\circ$ &  & meV & cm$^2$s$^{-1}$ $\times10^{-3}$ \\
  \hline
  21.79 & Classical & 160 & 0.042 \\
  21.79 & Quantum & 120 & 0.196 \\
  38.21 & Classical & 200 & 0.009 \\
  38.21 & Quantum & 120 & 0.201 \\
  \hline
 \end{tabular}
\end{table}

\begin{figure}[ht!]
\centering
 \includegraphics[width=1\linewidth]{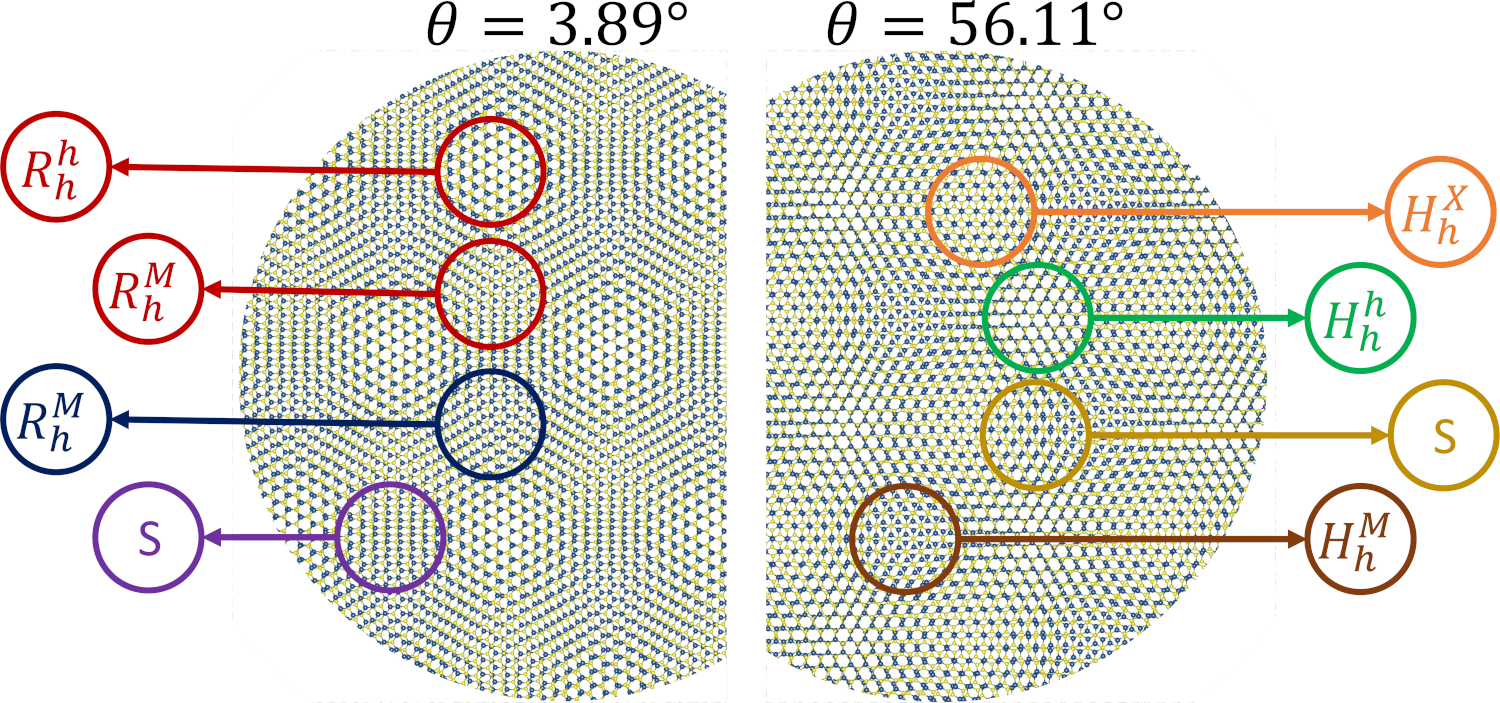} 
 \caption{\label{fig6} 
Superlattice structures in twisted bilayer MoS$_2$ with $\theta=3.89^\circ$ and $56.11^\circ$. The colored circle areas denote distinct high-symmetry local stacking regions ($H_h^h$, $H_h^X$, $R_h^M$ - domains, $H_h^M$ and $R_h^h$ - nodes) and the lower-symmetry solitons (S) in which H atoms were initially placed during metadynamics simulations. 
}
\end{figure}

In contrast, the low-angle twisted bilayers ($\theta = 3.89^\circ$, $R$-type stackings and $\theta = 56.11^\circ$, $H$-type stackings) develop pronounced domains, nodes, and soliton networks,\cite{arnold2023} resulting in strongly non-uniform diffusion landscapes with extended regions of high-symmetry stacking (Fig.~\ref{fig6}).
The nodes correspond to energetically unfavorable eclipsed configurations, namely $R_h^H$ for twist angles approaching $0^\circ$ and $H_h^M$ for twist angles approaching $60^\circ$.\cite{arnold2023}
These regions are characterized by larger interlayer separations and occupy only small spatial areas within the moiré supercell. 
In contrast, the domains are formed by energetically favorable staggered stackings, such as $R_h^M$ for small-angle $R$-type systems and $H_h^h$ and $H_h^X$ for twist angles approaching $60^\circ$.
These configurations possess smaller interlayer distances and occupy the largest fraction of the moiré pattern.
Separating neighboring domains are spatially extended soliton regions composed of continuously varying intermediate stackings. 
Since the isolated high-symmetry stackings were already analyzed in the previous sections, here we focus on the spatial dependence of hydrogen diffusion across these distinct moiré building blocks.

To probe the local diffusion landscape, WTMetaD simulations were initiated with hydrogen positioned in different regions of the twisted bilayers (see Fig.~\ref{fig6}).
Although lower $\Delta F$ might be expected for the predominantly $R$-type TBL-MoS$_2^{3.89}$ system, the calculated average barriers remain substantially larger than those obtained for isolated high-symmetry stackings.
Interestingly, the $H$-type TBL-MoS$_2^{56.11}$ system exhibits lower average diffusion barriers (of approximately 150~meV) compared to the $R$-type TBL-MoS$_2^{3.89}$ system (of approximately 197~meV), resulting in diffusion coefficients that are roughly three times larger (see Tab.~\ref{tab:table6}).
This behavior likely originates from the presence of the metastable $R_h^H$ node configuration in the $\theta = 3.89^\circ$ structure, which could not be stabilized in isolated high-symmetry simulations due to spontaneous relaxation toward lower-energy stackings.
In addition, the $\theta = 56.11^\circ$ system contains $H_h^X$ domains that significantly facilitate hydrogen diffusion, even though these regions occupy smaller spatial areas than $H_h^h$.
Note that our MLIP underestimates $\Delta F$ for this stacking, which also adds to the overall lower diffusion barrier for $\theta = 56.11^\circ$ system.
Additionally, the previously discussed high-symmetry stackings were investigated using bulk structures, while twisted systems are bilayer models.
It should also be noted that stronger layer shift against each other was observed during WTMetaD, which might be a technical artifact of the simulation.
The observed diffusion patters should, however, not be affected.

\begin{table}[ht!]
\centering
 \caption{\label{tab:table6} Results of hydrogen diffusion simulations in selected twisted bilayer MoS$_2$. Twist angles, $\theta$, classical free energy barriers, $\Delta F$, obtained from different starting positions of the H atom: in the node, $\Delta F_N$; in the domain, $\Delta F_D$, and in the soliton, $\Delta F_S$. The averaged classical free energy barriers, $\Delta \bar{F_C}$ are given together with the resulting averaged self-diffusion coefficients, $\bar{D_C}$. Additionally, the quantum free energy barrier, $\Delta F_Q$, and the corresponding self-diffusion coefficient, $D_Q$, were calculated with H atom starting position at the most favorable classical position, $\Delta F_D$.
 }
 \begin{tabular}{cccccc|cc}
 \\
 \hline
 \\[-0.8em]
  $\theta$ & $\Delta F_N$ & $\Delta F_D$ & $\Delta F_S$ & $\Delta \bar{F_C}$ & $\bar{D_C}$ & $\Delta {F_Q}$ & $D_Q$ \\
  $^\circ$ & meV & meV & meV & meV & {cm$^2$s$^{-1}$ $\times10^{-3}$} & meV & {cm$^2$s$^{-1}$ $\times10^{-3}$}\\
  \hline
  3.89 & 190 & 160 & 240 & 197 & 0.021 & 100 & 0.480\\
  56.11 & 155 & 135 & 160 & 150 & 0.074 & 45  & 3.941\\
  \hline
 \end{tabular}
\end{table}

Nevertheless, both low-angle twisted bilayers exhibit substantially larger diffusion barriers and correspondingly lower diffusivities than the isolated high-symmetry phases. 
Hydrogen preferentially resides within staggered stacking regions, where S atoms from neighboring layers provide a favorable local coordination environment and reduced diffusion barriers.
In contrast, the soliton regions introduce less favorable intermediate stackings that locally hinder H migration and increase the effective diffusion barriers across the moiré superlattice.
These findings demonstrate that hydrogen transport in twisted MoS$_2$ is governed by a delicate interplay between local stacking order, interlayer separation, and moiré-induced structural reconstruction.

Finally, we investigated NQEs in the reconstructed low-angle twisted bilayers, TBL-MoS$_2^{3.89}$ and TBL-MoS$_2^{56.11}$.
Based on the classical WTMetaD analysis discussed above, the PIMD simulations were initialized within the most favorable domain region obtained in the classical simulations, $\Delta F_D$, and the corresponding quantum free energy barriers ($\Delta F_Q$) were included in Table~\ref{tab:table6}.
NQEs substantially enhance hydrogen diffusion in both systems: $\Delta F$ decreases from 160 to 100~meV for TBL-MoS$_2^{3.89}$ and from 135 to only 45~meV for TBL-MoS$_2^{56.11}$, accompanied by corresponding increases in the self-diffusion coefficients from $0.021$ to $0.480$ and from $0.074$ to $3.941 \times10^{-3}$~cm$^2$,s$^{-1}$, respectively.
The particularly strong quantum reduction of 95~meV in TBL-MoS$_2^{56.11}$ results in the lowest quantum free-energy barrier among all investigated twisted bilayers (Tables~\ref{tab:table5} and \ref{tab:table6}).
More generally, comparison across the studied TBLs suggests that NQEs have a stronger influence on hydrogen diffusion in the $H$-type than in the $R$-type moiré structures.

More broadly, these findings demonstrate that twist engineering offers a powerful route for controlling hydrogen transport and designing spatially selective diffusion channels in layered 2D materials.

\section{Conclusion}
IIn conclusion, we have presented a comprehensive atomistic investigation of hydrogen and deuterium diffusion in layered and twisted MoS$_2$ systems by combining machine-learning interatomic potentials with well-tempered metadynamics and path-integral molecular dynamics.
This integrated framework enabled the explicit inclusion of nuclear quantum effects (NQEs) while retaining the length and time scales required to study structurally complex two-dimensional systems.

Our results demonstrate that NQEs play a decisive role in hydrogen transport in MoS$_2$ even at ambient temperature. 
Across all investigated polytypes, quantum fluctuations substantially soften the free-energy landscape, leading to pronounced reductions in diffusion barriers and significant enhancements of the hydrogen self-diffusion coefficients compared to classical nuclei simulations.
The magnitude of the quantum enhancement strongly depends on the local stacking environment, with the largest quantum correction observed for the $H_h^h$ configuration.

We further revealed a pronounced kinetic isotope effect arising from the mass dependence of quantum diffusion.
While deuterium exhibits systematically larger diffusion barriers and lower diffusivities than hydrogen already within the classical description, the inclusion of NQEs significantly amplifies these differences.
In particular, the quantum free-energy barriers of hydrogen and deuterium differ by approximately 31~meV, providing direct atomistic insight into isotope-selective transport mechanisms in layered materials.

Finally, our study of twisted bilayer MoS$_2$ demonstrates that moiré reconstruction generates diffusion landscapes with strong spatial variation.
Hydrogen transport is most favorable within staggered stacking domains and strongly suppressed in soliton regions and eclipsed configurations, producing local free-energy variations of up to about 100~meV across the moiré superlattice.
These findings establish a direct connection between local stacking order and hydrogen mobility, highlighting twist engineering as a promising strategy for controlling diffusion pathways in 2D materials.

Beyond providing fundamental insight into hydrogen diffusion and isotope effects in MoS$_2$, the presented MLIP-PIMD framework offers a powerful and transferable approach for investigating quantum transport phenomena in complex van der Waals (hetero)structures and related materials relevant for hydrogen storage, isotope separation, and clean-energy technologies.



All data related to this work will be available at https://doi.org/10.5281/zenodo.20007855 after the embargo is lifted on 01.09.2027.

\textbf{Acknowledgements} \par

This research was supported by the Deutsche Forschungsgemeinschaft (projects GRK 2721/1 and SFB 1415).
I.E. thanks Paul Philipp Wellmann for their technical support and introduction to the MACE code.

The authors gratefully acknowledge the computing time made available to them on the high-performance computer Otus at the NHR Center Paderborn Center for Parallel Computing (PC2). This center is jointly supported by the Federal Ministry of Research, Technology and Space and the state governments participating in the National High-Performance Computing (NHR) joint funding program (www.nhr-verein.de/en/our-partners).
The authors also acknowledge the high-performance computing center of ZIH Dresden and the Leipzig University Computing Centre.

Gemini and ChatGPT were used to improve the language and clarity of the manuscript. All calculations and data analyses were performed solely by the authors.

\textbf{Conflict of Interest}

The authors declare no conflict of interest.

\bibliography{references}

\section{Supporting Information}

\section{Methodology Details}
\subsection{Training setup}
In this study, we utilized the MACE (Message-Passing Atomic Cluster Expansion) machine learning model (version 0.3.13) to develop an efficient interatomic potential. The model was trained on a dataset that contained atomic structures with corresponding energy and force labels, divided randomly into training (70\%), validation (20\%), and test sets (10\%).

The training was performed on a CUDA-enabled device, using a training and validation batch size 8 and a maximum of 1000 epochs with last $\%$25 belongs to second stage. To ensure stability and prevent overfitting, an early stopping mechanism (20 steps) and exponential moving average (EMA) decay were applied. 
The loss function was weighted to prioritize accurate force predictions. Additionally, Stochastic Weight Averaging (SWA) was applied to enhance generalization. Weight of the first stage for energy and forces are set to 10 and 1000, respectively. The Adam optimizer with AMSGrad was used, with an initial learning rate of $1\times10^{-3}$ at first stage and $1\times10^{-6}$ at second stage, which was dynamically adjusted during training.

MACE employs spherical harmonics to represent 3D rotations, denoted as $Y_l^m$. 
In this work, the maximum order of the spherical harmonics was set to $l = 2$, 
while the maximum angular momentum was limited to $m = 1$. 

For improved accuracy, a cutoff radius of $6.0\,\text{\AA}$ was used. 
The model capacity was increased by setting the number of channels to 256. 
Additionally, the architecture consists of 2 message-passing layers, 
and the body-order correlation was chosen as 3.

\subsection{Generation and evaluation of MLIP}

Dataset construction is a crucial step in developing an MLIP. To describe hydrogen motion in MoS$_{2}$ reliably, the dataset must include diverse hydrogen environments and the relevant MoS$_2$ polytypes. Table~\ref{tab:table1} summarizes the composition of the structural dataset in terms of both polytype stacking and generation methodology. The largest contribution arises from adaptive learning (AL) iterations, which account for 599 structures out of a total of 944, indicating that the dataset is predominantly enriched through iterative sampling. Among the different stackings, $R^{M}_{h}$ and $H^{h}_{X}$ constitute the most extensively represented configurations with 308 and 285 structures, respectively, largely driven by their substantial AL contributions. In contrast, the $R^{h}_{h}$ stacking shows no participation from AL and remains comparatively limited with only 66 structures, primarily originating from cell optimization (Cell) and geometry optimization with hydrogen atom (H opt) calculations due to the fact that $R^{h}_{h}$ is a high-energy stacking and changes during the simulations to low-energy stacking. The 3R stacking exhibits a more balanced distribution across generation methods but contributes fewer total structures overall. The Cell, H opt, and Strain categories provide a relatively uniform baseline across stackings, with totals of 120, 135, and 90 structures, respectively, serving as the foundational dataset upon which AL further expands. Overall, the dataset reflects a strong emphasis on adaptive sampling while maintaining diversity across different structural generation approaches.

\begin{table}[ht!]
\centering
 \caption{\label{tab:table1} Breakdown of the structural dataset, categorized by polytype stacking and generation method: lattice relaxation (Cell), hydrogen optimization (H opt), strained pristine structures (Strain), and adaptive learning (AL) iterations.}
 \begin{tabular}{c|cccc|c}
 \hline
  Stacking & Cell  & H opt & Strain & AL & Total  \\
  \hline
  $H^{h}_{h}$ & 14 & 15 & 19 & 120  & 168  \\
  $H^{h}_{X}$ & 23 & 39 & 19 & 204  & 285  \\
  $R^{M}_{h}$ & 15 & 48 & 22 & 223  & 308  \\
  3R & 25 & 20 & 20 & 52  &  117 \\
  $R^{h}_{h}$ & 43 & 13 & 10 & 0  & 66  \\
\hline
  Total & 120 & 135 & 90 &  599 & 944  \\
  \hline
 \end{tabular}
\end{table}

Energies and forces for each structure and atomic references for H, S, and Mo atoms represented in the dataset (Table~\ref{tab:table2}) were calculated using density functional theory (DFT) as implemented in the CP2K software package~\cite{kuhne2020cp2k}.
The Quickstep module was employed, utilizing the Gaussian and Plane Waves (GPW) method. The exchange-correlation interactions were described using the r$^2$-SCAN meta-GGA functional, augmented with the rVV10 non-local van der Waals dispersion correction.\cite{grimme2021r2scan,ege6,rVV10_reference}
Core electrons were represented by Goedecker–Teter–Hutter (GTH)\cite{goedecker1996separable} pseudopotentials optimized using the UZH protocol (GTH-SCAN), and the valence wavefunctions were expanded using a triple-zeta valence basis set with polarization functions optimized for molecular systems (TZVP-MOLOPT-SCAN-GTH).
A plane-wave energy cutoff of 900 Ry was used for the 3R polytype, while a 2000 Ry cutoff was applied to the remaining polytypes.
A relative cutoff of 50 Ry was maintained for all systems. Calculations were performed on a $4 \times 4 \times 3$ supercell using a $\Gamma$-centered \textit{k}-point scheme. Multiplicity was set to 2 for H@MoS$_2$ structures.

\begin{table}[ht!]
\centering
 \caption{\label{tab:table2} Prediction errors for energy and atomic forces across the training, validation, and test subsets using MACE model. 
 }
 \begin{tabular}{lccc}
 \hline
  Set & RMSE Energy & RMSE Force & Relative Force \\
     & meV/atom & meV/\AA & $\%$ \\
  \hline
  Training & 3.2 & 9.3 & 1.70 \\
  Validation & 3.1 & 12.9 & 2.20 \\
  Test & 3.0 & 9.8 & 1.61 \\
  \hline
 \end{tabular}
\end{table}

\begin{figure*}[ht!]
 \includegraphics[width=1\textwidth]{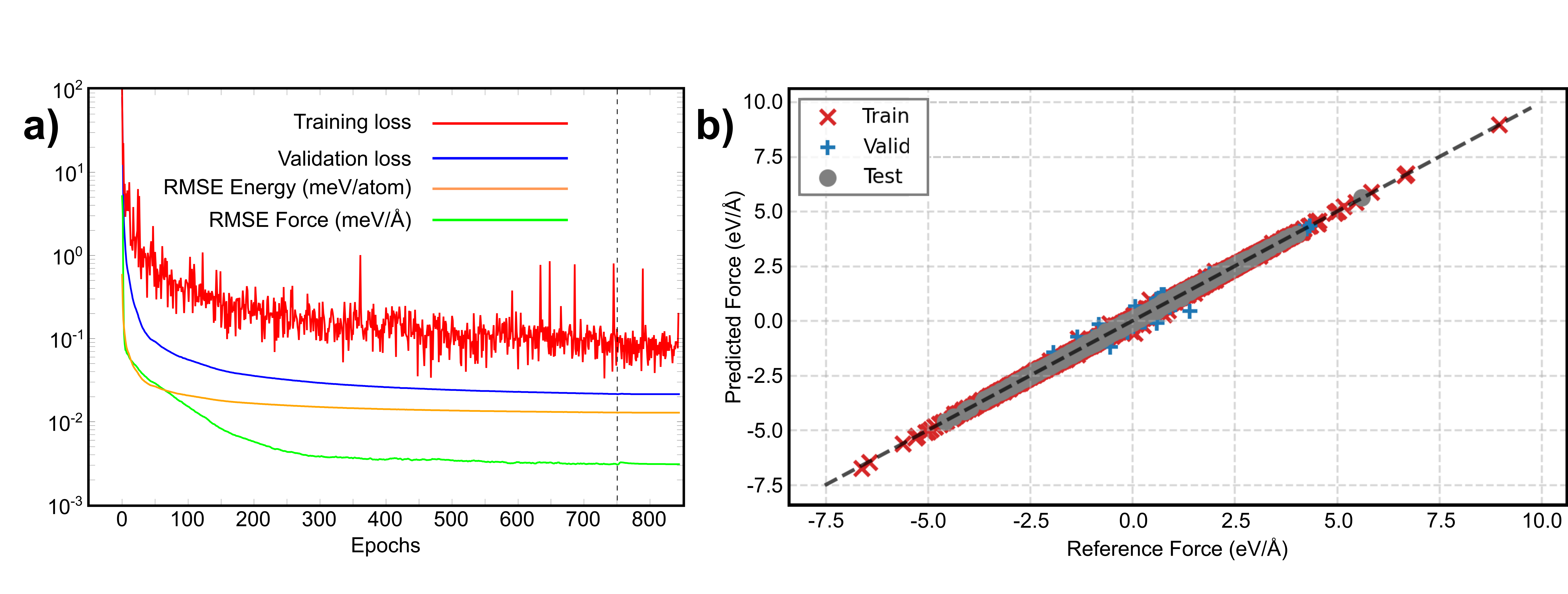}%
 \caption{\label{figsup6} 
 Training dynamics and predictive accuracy of the neural network potential. (a) Evolution of training and validation losses alongside energy and force RMSE over 850 epochs; the start of Stochastic Weight Averaging (SWA) is indicated by the vertical dashed line. (b) Parity plot of predicted versus reference atomic forces ($eV/{\text{\AA}}$) for the training, validation, and test subsets, demonstrating high correlation along the ideal $y=x$ dashed line.
 }
\end{figure*}

The predictive accuracy of the trained potential and its consistency across various structural configurations are quantitatively assessed in Table \ref{tab:table2}. A holistic evaluation of the results reveals a remarkably balanced error distribution between the training, validation, and test subsets. The evaluation of RMSE of training and validation throughout the machine learning procedure is also depicted in Fig.~\ref{figsup6}. 
The energy RMSE remains consistently low, with the test set achieving a value of 3.0 meV/atom. This high resolution is comparable to the inherent precision of first-principles methods, indicating that the model has successfully captured the global features of the potential energy surface. Furthermore, the minimal discrepancy between the training (3.2 meV/atom) and validation (3.1 meV/atom) energy errors demonstrates that the model is well-regularized and effectively avoids the pitfalls of overfitting, ensuring reliable energy predictions for unseen atomic environments.

In addition to scalar energy metrics, the vector-based force predictions provide deeper insight into the model's ability to describe the local curvature of the atomic landscape. The force RMSE values range between 9.3 and 12.9 meV/\AA, maintaining a relative force error of less than 2.20$\%$ across all partitions. Force errors are numerically higher than energy errors because forces represent the spatial derivatives of the energy and are thus more sensitive to local fluctuations.
The slight increase in force RMSE observed in the validation set (12.9 meV/\AA) likely reflects the inclusion of more diverse or high-strain configurations within that specific subset. However, the subsequent drop to 9.8 meV/\AA \quad in the test set confirms the model's robust generalization. Ultimately, the synergy between low energy residuals and precise force components validates the model’s readiness for complex structural optimizations and long-term thermodynamic simulations.

\section{Structural and Energetic Properties}

We compared lattice parameters and selected geometric components of all the high-symmetry stacking configurations using DFT and MLIP simulations (table \ref{tab:table3}). The errors are well below 1\%, indicating very good agreement between our MLIP and the DFT simulations.
The structural properties obtained from high-accuracy DFT calculations were described early. Geometry optimization was performed using the BFGS algorithm within the i-Pi framework, with a line search tolerance of $10^{-4}$ with MLIP. The structures were considered fully relaxed when the changes in total energy, maximum atomic force, and displacement simultaneously fell below $5 \times 10^{-6}$ a.u.

\begin{table}[ht!]
\centering
 \caption{\label{tab:table3} Comparision between MLIP and R2SCAN-rVV10 optimized structural properties of all investigated bulk stackings of MoS$_2$: lattice vectors,\textit{a}, \textit{b}, and \textit{c} (in \AA); bond lengths between metal and chalcogen atoms, $d_{S-Mo}$, and between chalcogen atoms within a single layer, $d_{S-S}$, (in \AA); interlayer distances measured between metal atoms, $d_{Mo-Mo}$, and between inner chalcogen atoms in the neighboring layers, $d_{I}$, (in \AA); and the bond lengths between chalcogen and hydrogen atoms, $d_{S-H}$, (in \AA). we used the structural definition as similar in our previous work.\cite{eren2024hydrogen}
 }
 \begin{tabular}{|l|c|ccccc|}
 \hline
    & Property & $H_{h}^{h}$ & $H_{h}^{X}$ & $R_{h}^{M}$ & 3R  &  $R_{h}^{h}$ \\
  \hline
  \hline
  \multirow{7}{*}{DFT} & \textit{a} = \textit{b} (\AA) & 3.164 & 3.169 & 3.167 & 3.167 & 3.162  \\
  
   & \textit{c} (\AA) & 12.145 & 12.195 & 12.081 & 18.111 & 13.244 \\
  
   & $d_{S-Mo}$ (\AA)  & 2.406 & 2.406 &  2.406 & 2.406 & 2.406 \\
  
   & $d_{S-S}$ (\AA) & 3.132 & 3.126 & 3.128 & 3.130 & 3.133  \\
  
   & $d_{Mo-Mo}$ (\AA) & 6.073 & 6.098 & 6.040 & 6.036 & 6.623  \\
  \
   & $d_{I}$ (\AA) & 2.952 & 2.968 & 2.906 & 2.886 & 3.489 \\
  
   & $d_{S-H}$ (\AA)  & 1.410 & 1.438 & 1.405 & 1.535 & 1.408 \\
  \hline
  \multirow{7}{*}{MLIP} & \textit{a} = \textit{b} (\AA) & 3.164 & 3.169 & 3.167 & 3.167 & 3.162 \\
  
   & \textit{c} (\AA) & 12.145 & 12.195 & 12.081 & 18.110 & 13.245 \\
  
   & $d_{S-Mo}$ (\AA)  & 2.406 & 2.406 &  2.406 &  2.406 &  2.407 \\
  
   & $d_{S-S}$ (\AA) & 3.133 & 3.127 & 3.127 & 3.128 & 3.118  \\
  
   & $d_{Mo-Mo}$ (\AA) & 6.073 & 6.098 & 6.041 & 6.037 & 6.623  \\
  
   & $d_{I}$ (\AA) & 2.935 & 2.966 & 2.909 & 2.906 & 3.370 \\
  
   & $d_{S-H}$ (\AA)  & 1.411 & 1.440 & 1.440 & 1.442 & 1.412 \\
  \hline
  \hline
  \multirow{7}{*}{Error} & \textit{a} = \textit{b} (\%) & 0 & 0 & 0 & 0 & 0 \\
  
   & \textit{c} (\%) & 0 & 0 & 0 & 0.006 & 0.008 \\
  
   & $d_{S-Mo}$ (\%)  & 0 & 0 & 0  & 0 & 0.042 \\
  
   & $d_{S-S}$ (\%) & 0.032 & 0.032 & 0.032 & 0.064 & 0.481  \\
  
   & $d_{Mo-Mo}$ (\%) & 0 & 0 & 0.017 & 0.017 & 0  \\
  
   & $d_{I}$ (\%) & 0.575 & 0.067 & 0.103 & 0.688 & 3.411 \\
  
   & $d_{S-H}$ (\%)  & 0.071 & 0.139 & 2.491 & 6.058 & 0.284 \\
  \hline
  \hline
 \end{tabular}
\end{table}

\begin{figure}[ht!]
\centering
 \includegraphics[width=0.4\textwidth]{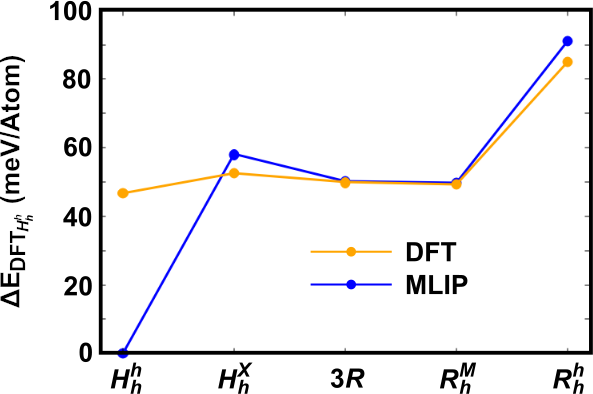}%
 \caption{\label{figsup4} 
 Validation of the MLIP against DFT for various polytype stackings. Relative energy differences per atom, where the $H_h^h$ stacking energy from DFT is taken as the reference ground state ($\Delta E = 0$).
 }
\end{figure}

Additionally, we calculated the relative energies per atom of all high-symmetry stackings with respect to the lowest-energy configuration as shown in Fig.~\ref{figsup4}. The stability of $H_h^h$ stacking is somewhat overestimated by our MLIP. 

\section{Vibrational Properties}

\begin{figure*}[ht!]
 \includegraphics[width=1\textwidth]{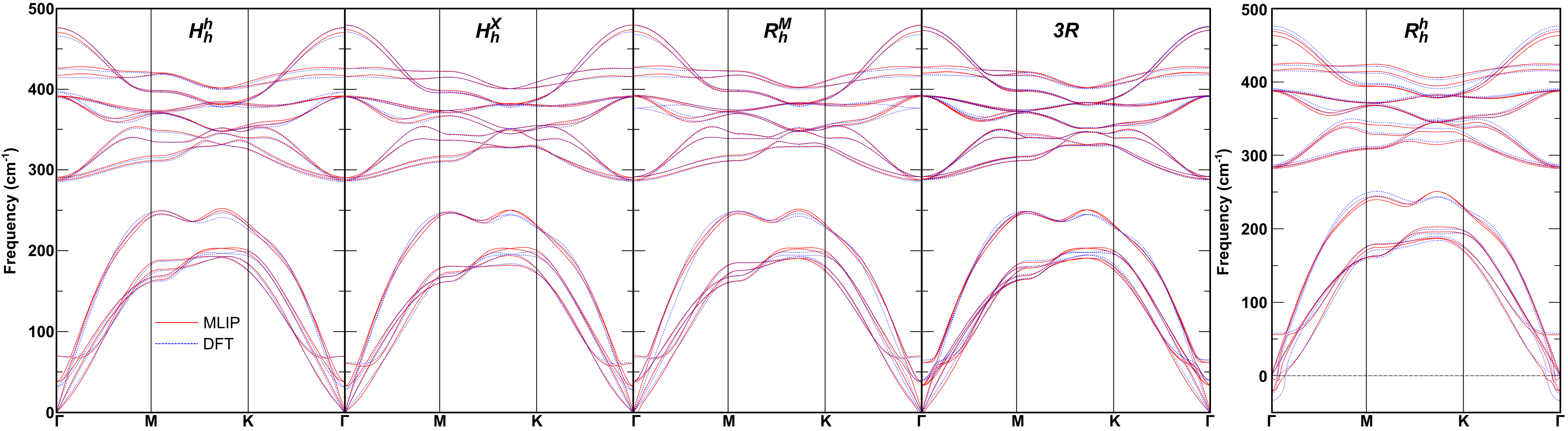}%
 \caption{\label{figsup5} 
 Comparison of the phonon dispersion relations calculated using Density Functional Theory (DFT) and the developed Machine Learning Interatomic Potential (MLIP) for bulk MoS$_2$ polytypes. The MLIP  are represented by red solid lines, while the DFT reference data are shown as blue dotted lines. The frequencies are given in $\text{cm}^{-1}$ along the $\Gamma$-M-K-$\Gamma$ high-symmetry path in the Brillouin zone.
 }
\end{figure*}

Next, phonon band structures were simulated using DFT and MLIP and compared for all the high-symmetry stackings in Fig.~\ref{figsup5} showing a very good agreement between the two levels of theory.

The phonon dispersions computed with the ML potential in Fig. \ref{figsup5} show a clear separation between stacking-insensitive intralayer vibrations and stacking-sensitive interlayer dynamics. The high-frequency optical manifolds are almost superposed for $H_{h}^{h}$, $H_{h}^{X}$, 3R, $R_{h}^{M}$, and $R_{h}^{h}$, with the main $\Gamma$-point groups centered near approximately 287, 388–411, and 470 cm$^{-1}$, indicating that the potential preserves the dominant intralayer Mo–S force constants across all five registries. By contrast, the largest registry dependence appears in the low-energy sector, particularly in the interlayer shear- and breathing-like modes and the adjacent low-frequency optical branches below roughly 50–60 cm$^{-1}$ and, more broadly, below about 200 cm$^{-1}$, where the restoring forces are governed primarily by weak interlayer coupling.  No branch crosses below zero anywhere along $\Gamma$–M–K–$\Gamma$, so none of the five stackings shows an obvious harmonic instability on the plotted path. This hierarchy is exactly what is expected for layered MoS$_2$; prior Raman and first-principles studies show that low-frequency interlayer modes are strongly stacking dependent, whereas the high-frequency intralayer modes change only weakly between 2H- and 3R-type crystals.\cite{lee2016raman}

Comparison with literature shows that the ML phonon spectra fall within the established MoS$_2$ frequency windows and reproduces the accepted stacking trends. The most direct benchmark is the 2H phase. Molina-Sánchez and Wirtz reported bulk 2H-MoS$_2$ $\Gamma$-point frequencies of 35.2 cm$^{-1}$ for the shear E$_{2g}^2$ mode, 288.7 cm$^{-1}$ for E$_{1g}$, 387.8 cm$^{-1}$ for E$_{2g}^1$, 391.2 cm$^{-1}$ for E$_{1u}$, 412.0 cm$^{-1}$ for A$_{1g}$, and 469.4 cm$^{-1}$ for A$_{2u}$, with an E$_{1u}$ LO–TO splitting of 2.8 cm$^{-1}$.\cite{molina2011phonons} Ataca \textit{et al}. obtained closely related 2H values from GGA+D calculations using experimental lattice constants, namely 277.8 cm$^{-1}$ for E$_{1g}$, 381.3 cm$^{-1}$ for E$_{2g}^1$, and 407.7 cm$^{-1}$ for A$_{1g}$; the same study also predicted A$_{1g}^{'}$ frequencies of 404.9 cm$^{-1}$ for bilayer and 405.9 cm$^{-1}$ for trilayer MoS$_2$.\cite{ataca2011comparative}
The ML branches in Fig. \ref{figsup5} occupy these same optical windows, with the upper optical manifolds clustering near the established E$_{1g}$, E$_{2g}^1$/E$_{1u}$, and A$_{1g}$/A$_{2u}$ regions rather than deviating into unphysical frequency ranges. At the same time, the strongest stacking dependence in the ML spectra is concentrated in the low-frequency sector, whereas the high-frequency intralayer modes remain comparatively insensitive to registry, which is exactly the trend expected for layered MoS$_2$. Molina-Sánchez and Wirtz further showed that the A$_{1g}$ mode increases with increasing layer number, while the E$_{2g}^1$ mode decreases, and attributed the latter trend to enhanced dielectric screening of the long-range Coulomb interaction in multilayer and bulk MoS$_2$.\cite{molina2011phonons}

\section{Dynamical Properties}

\begin{figure*}[ht!]
\centering
 \includegraphics[width=1\textwidth]{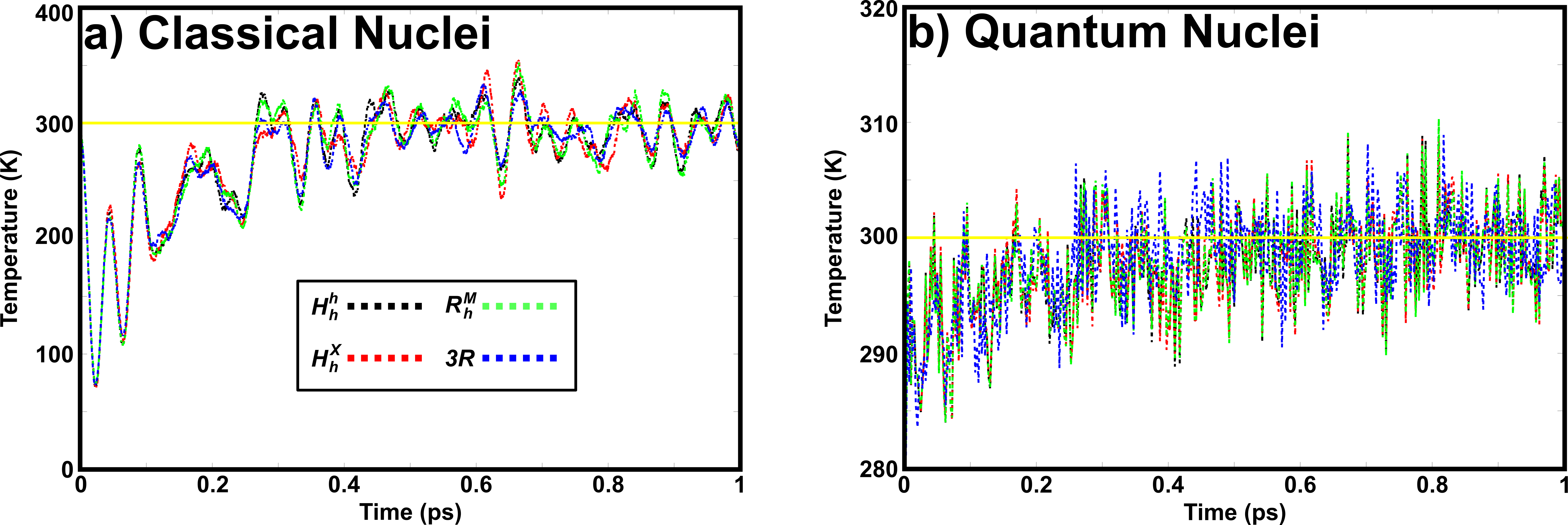}%
 \caption{\label{fig1} 
 Thermal equilibration and dynamical stability of MoS$_2$ polytypes. (a) Instantaneous temperature as function of time during classical MD simulations. (b) Temperature evolution during PIMD simulations, accounting for NQE. Both panels illustrate the convergence toward the target temperature of 300 K (yellow line) within 1 ps, demonstrating the robustness of the MLIP. Note that MD utilizes the SVR thermostat, while PIMD employs the PILE-G thermostat.
 }
\end{figure*}

The stabilization of temperature before the initialization of metadynamics is a critically important step to ensure that subsequent free energy calculations are sampled from a well-equilibrated phase space. As detailed in the methodology section, the temperature evolution for the MoS$_2$ polytypes is monitored to verify that the machine learning interatomic potential maintains thermodynamic consistency. As shown in Fig.~\ref{fig1}a, the classical molecular dynamics simulations utilizing the stochastic velocity rescaling thermostat exhibit rapid convergence to the target temperature of 300 K. The instantaneous temperature fluctuations for the $H_h^h$, $H_h^X$, $R_h^M$, and $3R$ phases stabilize within the first picosecond of the trajectory. This rapid equilibration confirms that the potential energy surface is well-represented and that the drive toward thermodynamic equilibrium is numerically consistent for all considered structural configurations.
The robustness of the potential is further validated through path integral molecular dynamics simulations, which incorporate nuclear quantum effects via the PILE-G thermostat. Fig.~\ref{fig1}b illustrates that the system maintains a stable temperature profile centered at 300 K even when accounting for the quantized nature of the nuclei. Although the quantum trajectories exhibit a different fluctuation frequency compared to the classical case, the overall convergence remains tightly bound to the equilibrium line. The ability of the potential to handle both classical and quantum ensembles without nonphysical drifts or instabilities signifies its reliability for long-term production runs and complex phase stability studies in MoS$_2$ systems.

\section{Convergence of the number of the beads}

\begin{figure*}[h!]
\centering
 \includegraphics[width=0.8\textwidth]{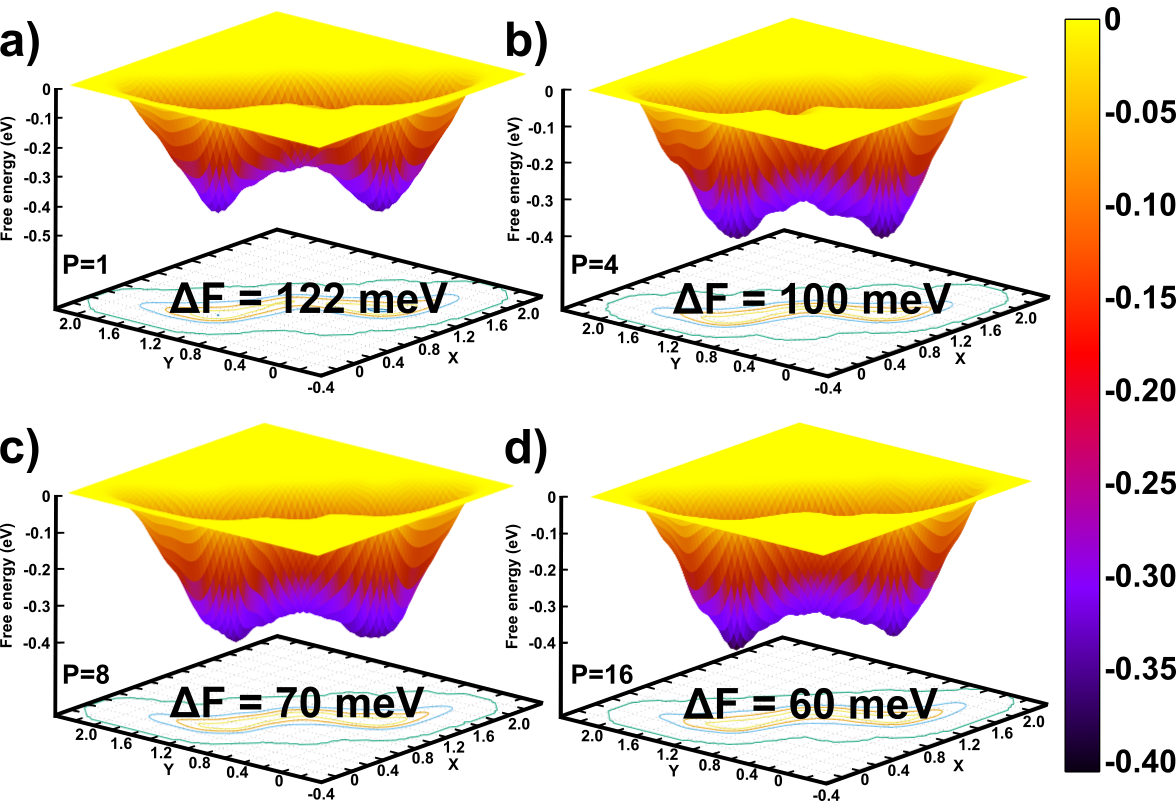}%
 \caption{\label{figsup3} 
 Convergence of number of beads via well-tempered PIMD calculations for H atom in $H_h^h$ stacking.
 }
\end{figure*}

The convergence of number of beads via well-tempered PIMD calculations for H atom in $H_h^h$ stacking is shown in Fig.~\ref{figsup3}.

\end{document}